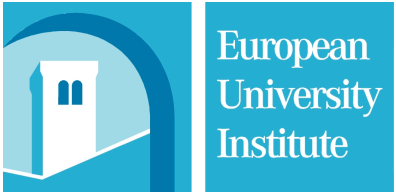

# European Journal of Legal Studies

**Abstract:**

*Over a century ago, Oliver Wendell Holmes invited scholars to look at law through the lens of probability theory: 'The prophecies of what the courts will do in fact, and nothing more pretentious, are what I mean by the law'. But Holmes himself, and few others, have taken up this intriguing invitation. As such, in place of previous approaches to the study of law, this paper presents a non-normative, mathematical approach to law and the legal process. Specifically, we present a formal Bayesian model of civil and criminal litigation, or what we refer to as the 'litigation game'; that is, instead of focusing on the rules of civil or criminal procedure or substantive legal doctrine, we ask and attempt to answer a mathematical question: what is the posterior probability that a defendant in a civil or criminal trial will be found liable, given that the defendant has, in fact, committed a wrongful act?*

# ESSAY

## A BAYESIAN MODEL OF THE LITIGATION GAME

### F.E. GUERRA-PUJOL[*]

TABLE OF CONTENTS

[*] Associate Professor of Law, Barry University Dwayne O. Andreas School of Law, e-mail: eguerra@mail.barry.edu. The author thanks Dean Leticia Diaz and the Barry University for awarding him a summer grant in support of the research presented in this paper. The author also wishes to thank his colleague and close friend Orlando Ivan Martinez-Garcia for pointing him in the direction of mathematics so many years ago. Without Mr Martinez-Garcia's intellectual and personal friendship, the author would still find himself metaphorically lost at sea, clinging to the flotsam of traditional legal scholarship. Lastly, the author thanks Sydjia Robinson for her research assistance. In addition, Ms Robinson patiently listened to the author's ideas and made many useful suggestions during the many weeks spent researching and writing this paper.

## 1. Introduction

Why do mathematics and legal studies travel in such different directions; why is it that mathematicians and lawyers rarely take the time to speak to one another? Mathematics is based on axioms and abstract symbols, beautiful patterns and elegant proofs, while law has traditionally been a linguistic game, one based on semantics, simple syllogisms, and reasoning by analogy.[1] This paper, however, attempts to bridge the gap between these apparently disparate disciplines by looking at the process of litigation through the lens of probability theory.

It was over a century ago that Oliver Wendell Holmes first invited scholars to look at the law through this lens: 'The prophecies of what the courts will do in fact, and nothing more pretentious, are what I mean by the law'.[2] But Holmes himself and few other scholars have taken up this intriguing invitation. As such, in place of previous approaches to the study of law, this paper presents a non-normative, mathematical approach to law and the legal process. Specifically, we turn to Thomas Bayes, not William Blackstone, for inspiration and present a formal Bayesian model of civil and criminal litigation, or what we refer to as the 'litigation game'.[3] That is, instead of focusing on the rules of civil or criminal procedure or substantive legal doctrine, we ask and attempt to answer a mathematical question: what is the posterior probability that a defendant in a civil or criminal trial will be found liable, given that the defendant has, in fact, committed a wrongful act?

---

[1] Compare, for example, the overview of mathematics in Keith Devlin, *Mathematics: The Science of Patterns* (Holt 1994) with the description of the methods of legal reasoning in Edward H. Levi, *An Introduction to Legal Reasoning* (UChicago 1949).

[2] Oliver Wendell Holmes, 'The Path of the Law' (1897) 10 Harvard LR 457, 461.

[3] The main reason we refer to the process of adjudication or litigation as a game is to emphasize the interdependence of litigation outcomes. In summary, the outcome of a civil or criminal trial depends not only on the guilt or innocence of the defendant but also on the strategic moves made by the parties. For a vivid presentation of the idea of interdependence and a summary of strategic ploys, see Thomas C. Schelling, *The Strategy of Conflict* (Harvard 1960). To our knowledge, the first use of the term 'litigation game' appears in Marc Galanter, 'Why the "Haves" Come Out Ahead' (1974) 9 LSR 95, reprinted in David Kennedy and William W. Fisher, *The Canon of American Legal Thought* (Princeton 2006) 495-545.

The remainder of this paper is organized as follows: following this brief introduction, Section 2 briefly summarizes previous approaches to the study of law: legal formalism, legal realism, and economic analysis of law. Next, Section 3 presents Bayes' rule of conditional probability and explains the logic of the Bayesian or probabilistic approach to litigation, while Section 4 presents our formal Bayesian model of the process of adjudication, the litigation game. Section 5 concludes with a confession by the author.

## 2. Brief summary of previous approaches to the study of law

Since the classical days of Christopher Columbus Langdell, Anglo-American scholars have produced three important intellectual movements, three Kuhnian 'paradigm-shifts'[4] often referred to as legal formalism, legal realism, and economic analysis of law (or 'law and economics').[5] Beginning with Dean Langdell, the so-called 'legal formalists' presented law and the legal system as a rational and self-contained logical system.[6] Then came Oliver Wendell Holmes and the more radical 'legal realists', who, broadly speaking, saw law as a form of politics.[7] Where the formalists saw coherence and logical syllogisms, the realists saw politics and radical indeterminacy. But the realists and formalists shared the same fundamental flaw: they were unable to offer a workable and forward-looking research agenda. The law-and-economics movement thus attempted to fill this academic void, although some scholars have persuasively argued that economic analysis as applied to law is just another form of legal formalism.[8]

---

[4] The influential idea of scientific revolutions (or 'paradigm-shifts') is set forth in Thomas S. Kuhn, *The Structure of Scientific Revolutions* (3rd edn, UChicago 1996).

[5] A general history of previous legal scholarship is more fully presented in Kennedy & Fisher, *The Canon of American Legal Thought* (n 3) 1-12. See also Pierre Schlag, 'Spam Jurisprudence' (2009) 97 Georgetown LJ 803, 821; Richard A. Posner, 'The State of Legal Scholarship Today' (2009) 97 Georgetown LJ 845, 847.

[6] This formalist view of law appears in the preface to Dean Langdell's famous casebook, Christopher Columbus Langdell, *Selection of Cases on the Law of Contracts* (Little Brown 1871) v-vii.

Nevertheless, economic analysis of law not only offered a forward-looking research program for legal studies, economists also imported another important innovation to legal scholarship: the use of mathematics and mathematical methods in law. Economists, not lawyers nor mathematicians, thus played a leading role in systematically applying mathematical methods to law. Perhaps the most celebrated use of mathematics in legal studies is found in the opening pages of Ronald Coase's landmark paper, 'The Problem of Social Cost', in which Professor Coase presents a simple and straightforward arithmetical table to illustrate the reciprocal nature of negative externalities.[9] Following Professor Coase's famous arithmetical analysis of the problem of harmful effects, many economists, and even some legal scholars, have continued to apply ever-more sophisticated mathematical methods to legal problems.[10]

---

[7] Aside from Holmes, 'The Path of the Law' (n 2), another well-known statement of this position appears in Judge William Andrews's dissenting opinion in *Palsgraf v Long Island RR Co* [1928] 248 NY 339, 162 NE 99 (Andrews J). See also Felix Cohen, 'Transcendental Nonsense and the Functional Approach' (1935) 35 Columbia LR 809.

[8] See, for example, Arthur Alan Leff, 'Economic Analysis of Law: Some Realism about Nominalism' (1976) 60 Virginia LR 451.

[9] Ronald H. Coase, 'The Problem of Social Cost' (1960) 3 JLE 1, 3.

[10] For example, one of the leading proponents of the use of sophisticated mathematical models in law is the economist Gary Becker, who has applied such methods to illuminate a wide range of legal fields, including criminal law, employment discrimination, and even family law. Gary S. Becker, *The Economic Approach to Human Behavior* (UChicago 1976).

For many scholars, however, the Achilles' heel of the economic approach to law is the 'rational actor model' of human behavior, the standard assumption of rationality. Broadly speaking, the law-and-economics literature tends to assume that legal actors have perfection information and are able to measure and weigh the benefits and costs of their actions, that is, that they are rational calculators of the expected utility of their decisions.[11] In contrast, in this paper we abandon the rationality assumption through the use of Bayesian analysis. Furthermore, in place of previous approaches to the study of law, such as legal formalism and legal realism, we present a formal mathematical model of civil and criminal litigation. Before presenting our Bayesian model of the litigation game, we briefly explain the logic of Bayesian reasoning below.

## 3. The Bayesian approach to litigation

In contrast to previous approaches to legal studies, our approach is Bayesian or probabilistic, since our model of the litigation game is derived from Bayes' theorem or Bayes' rule of conditional probability.[12] In summary, Bayes' theorem can be expressed in algebraic terms as follows:

---

[11] In defense of the rational actor model, it is worth noting that economists generally assume that legal and other actors maximize their utility functions, not because this is a realistic assumption (it is not), but rather to apply the methods of calculus and make economic analysis of legal problems mathematically tractable or, in the words of one writer, 'soluble'. Peter Medawar, *The Art of the Soluble* (Methuen 1967) 7.

[12] The first description of Bayes' ideas appears in Thomas Bayes, 'An Essay Towards Solving a Problem in the Doctrine of Chances' (1763) 53 Philosophical Transactions of the Royal Society of London 370. In addition, a comprehensive and useful survey of Bayes' theorem appears in James Joyce, 'Bayes' theorem', *The Stanford Encyclopedia of Philosophy* (Fall edn, 2008) <http://plato.stanford.edu/archives/fall2008/entries/bayes-theorem/> accessed 19 November 2011. In addition, a highly readable and simplified summary of Bayes' contributions to probability theory may be found in Brian Everitt, *Chance Rules* (Springer 2008) 88-104.

$$Pr(A|B) = ([Pr(B|A)] \times [Pr(A)]) \div Pr(B)$$

Explained in words, Bayes's formidable-looking formula may be broken down into the following five parts:

(i) The term on the left-hand side of the equation, $Pr(A|B)$, refers to the conditional probability (or posterior probability) of event A, given the occurrence of event B.

(ii) The right-hand side of the equation is a fraction: the numerator contains two parts, $Pr(B|A) \times Pr(A)$, while the denominator consists of one term, $Pr(B)$.

(iii) The first term in the numerator, $Pr(B|A)$, refers to the conditional probability of event B, given the occurrence of event A.

(iv) The second term in numerator, $Pr(A)$, refers to the prior probability (or unconditional probability) of event A, that is, the probability of A in the absence of any information about event B.

(v) Lastly, the denominator, $Pr(B)$, is the prior probability (or unconditional probability) of event B in the absence of any information about event A.

In the remainder of this paper, we will equate the term 'guilty' (or the letter 'A') with the event that the defendant in a particular litigation game has committed a wrongful or unlawful act, that is, an act for which he should be civilly or criminally liable.[13] In addition, we will equate the term the symbol + (or the letter 'B') with the event that the defendant is actually found liable at trial for the commission of a civil or criminal wrongful act.[14] In other words, B or + is the probability of a positive litigation outcome from the perspective of the moving party in the litigation game, the plaintiff (in a civil trial) or the prosecutor (in a criminal trial). In other words, the main idea here is that the moving party – the plaintiff or prosecutor, as the case may be – obtains a favorable or positive outcome, which is denoted by the symbol +, when the defendant is found civilly or criminally liable at trial. Our Bayesian model of the litigation game thus poses the following fundamental question: what is the posterior probability that a defendant in a civil or criminal trial will be found liable, given that the defendant has not, in fact, committed any wrongful act?[15]

---

[13] The term $Pr(A)$ or $Pr(guilty)$ (in contrast to the terms 'A' or 'guilty') refers to the prior probability in the absence of additional information that this event (i.e., the imposition of civil or criminal liability) has in fact occurred.

At this point, we must introduce and formally define the technical concepts of 'sensitivity' and 'specificity'. In the context of our Bayesian model of the litigation game, these concepts refer to the underlying reliability of a civil or criminal trial to distinguish between guilty and innocent defendants. Since civil or criminal liability should be imposed only on guilty defendants, i.e., defendants who have in fact committed an unlawful wrongful act, sensitivity and specificity are thus important values. Specifically, the 'sensitivity' of the litigation game – written as $Pr(B|A)$ or, in our model, $Pr(+|guilty)$ – indicates how well a civil or criminal trial is able to correctly impose liability on guilty defendants. In summary, this measure is defined formally as the probability of a positive litigation outcome (i.e., liability imposed on the defendant, which represents a 'positive' outcome from the plaintiff's or prosecutor's perspective), given that the defendant being tried has actually committed an unlawful wrongful act.

---

[14] In other words, the symbol + and the term 'positive litigation outcome' is not meant to convey a pro-plaintiff or pro-prosecutor bias; instead, we use it to indicate a litigation outcome in which civil or criminal liability is imposed on the defendant.

[15] Like the term 'litigation', we define 'wrongful act' broadly to include both civil wrongs, such as torts and breaches of contract, as well as criminal wrongs, such as homicide and theft.

By contrast, the 'specificity' of the litigation game, which may be written as Pr( | innocent), reflects how well a civil or criminal trial is able to correctly screen out innocent defendants. This measure is defined formally as the probability of a negative litigation outcome (i.e., no liability imposed on the defendant, which represents a 'negative' outcome from the perspective of the moving party, plaintiff or prosecutor), given that the defendant has not committed a wrongful act.

Before presenting our Bayesian model in section 4 below, we wish to make three general points about Bayesian reasoning in general. First, the basic idea behind Bayes's theorem is the idea that the conditional probability of event A, such as a defendant being found liable, given the occurrence of another event B, the defendant's commission of a wrongful act, not only depends on the strength of the relationship between A and B; it also depends on the prior probability of each event. Thus, according to Bayes's theorem, the probability that a defendant in a civil action will be found liable (for tort, breach of contract, etc.), given that a plaintiff has brought an action against the defendant, will generally depend on two sets of probabilities: (i) the likelihood of the defendant being found liable given the strength of plaintiff's claim, and (ii) the prior probabilities or success rates of plaintiffs and defendants generally.

Secondly, notice that the probability of some event A conditional on some other event B is not the same as the conditional probability of event B given event A, or stated formally: $\Pr(A|B)$ is not equal to $\Pr(B|A)$.[16] For example, the probability that a defendant will be found civilly or criminally liable, given that the defendant has committed some wrongful act (the commission of a tort, a breach of contract, a crime, etc.), is not the same as the probability that the defendant's wrongful conduct will result in liability, given that the plaintiff brings an a civil or criminal action against the defendant. We will explore this idea further in section 4 below, when we present our Bayesian model of the litigation game.

---

[16] This point is also made in Everitt, *Chance Rules* (n 12) 90.

Lastly, it is also worth noting that our Bayesian model of the litigation game does not rely on any unrealistic assumptions about human rationality, nor does it require any detailed information about any particular rules of procedure or about substantive legal doctrine. Since such procedural rules and legal doctrines are often unclear, contested, and subject to manipulation,[17] one can begin to appreciate the advantage of the Bayesian approach to civil and criminal litigation. In place of hunches, verbal arguments, and the inevitable 'thrust and parry' of competing interpretations of indeterminate rules and doctrines,[18] our Bayesian approach to the litigation game attempts to understand the legal process from a probabilistic perspective.

## 4. The model

Here, we present a stylized Bayesian model of the litigation game. To do so, we make a number of simplifying assumptions about the litigation process. First, we define 'litigation' broadly to include both criminal and civil cases. In essence, the litigation game (whether civil or criminal) is a contest in which the moving party, the plaintiff or the prosecutor, attempts to impose civil or criminal liability on the defendant for the commission of an unlawful or wrongful act (whether civil or criminal in nature). And likewise, seen from the defendant's perspective, litigation is a contest in which defendants attempt to avoid the imposition of liability. Our model thus presents litigation as a game with two possible outcomes: (i) positive and (ii) negative (hence the term, 'litigation game'). Specifically, a positive outcome occurs when the moving party successfully imposes civil or criminal liability on the defendant; a negative outcome, when the defendant is able to avoid the imposition of liability.[19]

---

[17] See, for example, Gordon Tullock, *The Logic of the Law* (Basic Books 1971) 48-49.

[18] Karl Llewellyn, *The Common Law Tradition* (Little Brown 1960) 522-529.

Before proceeding, notice that the relevant rules of procedure (i.e., the rules of the litigation game) as well as the scope and legal meaning of 'wrongful acts' and the types of legal liability imposed on wrongful actors are not relevant and are thus extraneous to our simplified model. In place of traditional legal analysis, our model abstracts from the morass of legal materials and takes these features of the legal landscape as a given. Stated formally, these details are exogenous or external to our model. Having stated our simplifying assumptions, we now proceed to apply Bayes' theorem to the litigation process. Recall the statement of Bayes' rule from the previous section of this paper:

$$\Pr(A|B) = [\Pr(B|A) \times \Pr(A)] \div \Pr(B)$$

Translated into the language of our model of the litigation game, Bayes' rule may now be restated as follows:

$$\Pr(guilty|+) = [\Pr(+|guilty) \times \Pr(guilty)] \div \Pr(+)$$

In other words, we want to find the posterior probability, $\Pr(guilty|+)$, that a defendant will be found liable at trial, given that he or she has actually committed some wrongful act. Ideally, of course, liability should be imposed only when a defendant has actually committed a wrongful act, and conversely, no liability should be imposed on innocent defendants.[20] But in reality, false negatives and false positives will occur for a wide variety of reasons, such as heightened pleading standards and abuse of discovery in civil actions and prosecutorial discretion and prosecutorial misconduct in criminal cases.[21] Stated colloquially, some guilty defendants will be able to avoid the imposition of liability, while some innocent ones will be punished.

---

[19] As an aside, we note that our Bayesian model of the litigation game ignores the temporal dimension of adjudication ('time costs' and the problem of delay); instead, we assume for simplicity that litigation is an instantaneous event, like a coin toss or the roll of a die. For a deeper exploration of the problem of time scarcity, see F.E. Guerra-Pujol, 'Time Scarcity and the Problem of Social Replicants: Clones and the Coase Theorem' (forthcoming) 2 JLSD.

[20] In an ideal or perfect legal system, the value for $\Pr(guilty|+)$ should be equal to or close one. Stated formally, $\Pr(A|B) \approx 1$.

Our Bayesian approach to the litigation game takes into account both (i) the possibility of a false positive (i.e., the imposition of liability when the defendant has not committed any wrongful act) as well as (ii) the possibility a false negative (no liability even though the defendant has, in fact, committed a wrongful act). The purpose of our stylized model, however, is not to explore the many systemic imperfections procedural or practical or otherwise in the existing legal system, imperfections contributing to the problem of false positives and negatives. This well-worn path has been explored by many others.[22] Instead, the goal of our model is to solve for $\Pr(\text{guilty}|+)$ and answer the following key question: how reliable is the litigation game, that is, how likely is it that a defendant who is found liable is, in fact, actually guilty of committing a wrongful act?

We will consider four possible scenarios or types of litigation games in the remainder of this paper: (i) non-random adjudication with risk-averse or 'virtuous' moving parties, (ii) non-random adjudication with risk-loving or 'less-than-virtuous' moving parties, (iii), random adjudication with risk-averse moving parties, and (iv) random adjudication with risk-loving moving parties. This schema may thus be depicted in tabular form as follows:

---

[21] In the context of the litigation game, a false positive or Type I error occurs when a defendant who has not committed a wrongful act is nevertheless found liable for the commission of such act. By contrast, a false negative or Type II error occurs when a tortious or guilty defendant is able to avoid the imposition of liability.

[22] See, for example, Galanter, 'Why the "Haves" Come Out Ahead' (n 3); see also Tullock, *The Logic of Law* (n 17).

| Type of litigation game | Type of moving party |
|---|---|
| non-random adjudication | risk-averse |
| non-random adjudication | risk-loving |
| random adjudication | risk-averse |
| random adjudication | risk-loving |

In summary, the adjudication variable in our model refers to the reliability or screening effectiveness of the process of adjudication. Specifically, 'non-random adjudication' refers to litigation games that are 90% sensitive and 90% specific, an assumption based on the classic and oft-repeated legal maxim 'it is better that ten guilty men escape than that one innocent suffer'.[23] Random adjudication, in contrast to non-random adjudication, occurs when litigation games are only 50% sensitive and 50% specific and thus no more reliable than the toss of a coin.[24] As an aside, it is worth asking, why would the process of adjudication ever produce a 'random' outcome in the real world? One possibility is that the level of randomness or unpredictability of adjudication might be a function of the level of complexity or ambiguity of legal rules. Consider, for example, the 'reasonable man' standard in tort law: the more complex or 'open-textured' the rules of substantive and procedural law are, the more random the litigation game will be.[25] Also, before proceeding, notice that the adjudication variable can never be 100% sensitive nor 100% specific since errors are inevitable in any process of adjudication, regardless of the litigation procedures that are in place.

---

[23] 4 Bl Comm 358, quoted in Eugene Volokh, 'N Guilty Men' (1997) 146 U Penn LR 173.

[24] With respect to trials with two possible outcomes (e.g., positive and negative, or heads and tails), by definition a random outcome cannot occur with more nor with less than 50% probability. We thank our research assistant, Sydjia Robinson, for pointing out this observation to us.

[25] See HLA Hart, *The Concept of Law* (2nd edn, Clarendon Press 1994); see also Gordon Tullock, *The Logic of the Law* (n 17) 48-49. For further exploration of this topic, see F.E. Guerra-Pujol, 'Chance and Litigation' (forthcoming) 21 Boston U Public Interest LJ.

In addition, the term 'risk-averse' or 'virtuous', as applied to moving parties, refers to plaintiffs and prosecutors who play the litigation game only when they are at least 90% certain that the named defendant has committed an unlawful wrongful act, while 'risk-loving' or 'less-than-virtuous' moving parties refers to plaintiffs and prosecutors who are willing to play the litigation game even when they are only 60% certain that the named defendant has committed a wrongful act. Stated colloquially, virtuous plaintiffs are civil plaintiffs who rarely file frivolous claims and criminal prosecutors who rarely abuse their discretion; by contrast, less-than-virtuous moving parties are more willing to gamble on litigation games than their more virtuous colleagues.

## 4.1 Non-random adjudication with risk-averse moving parties

Suppose the litigation game is 90% sensitive and 90% specific, that is, suppose the process of litigation is able to determine correctly, at least 90% of the time, when a defendant has committed a wrongful act, and suppose further that the process will also determine correctly, again at least 90% of the time, when a defendant has not, in fact, committed a wrongful act. The intuition behind this assumption (non-random adjudication) is that reliable legal procedures will tend to produce just and fair results.[26] Of course, the existence of reliable adjudication procedures in which liability is imposed only on guilty defendants is not a sufficient condition for justice. When a defendant has broken an unjust or unfair law (licensure requirements and racial segregation laws quickly come to mind), justice would be better served by an unreliable adjudication procedure (i.e., by not enforcing the unjust or unfair law in the first place). But putting aside the underlying meaning of justice, such a litigation game appears to be a highly accurate one, since it will correctly determine with 90% probability, or nine times out of 10, whether the defendant has or has not committed a wrongful act, an essential precondition before liability may justly be imposed.

---

[26] Henry M. Hart and Albert M. Sacks, *The Legal Process* (William N. Eskridge and Philip P. Frickey eds, Foundation 1994).

Nevertheless, even in the absence of unjust laws, our model of non-random adjudication still suffers from a 10% error rate. Given this error rate, we must turn to Bayes' rule to determine the posterior probability that liability will nevertheless be incorrectly imposed on an innocent defendant, that is to say, the probability that a defendant who has not committed a wrongful act will be incorrectly classified as a wrongful or guilty defendant. To apply Bayes' theorem, we must find the prior probability that any given defendant, selected at random, has in fact committed a wrongful act. What is this prior probability?

First, let the term 'guilt' stand for a guilty defendant, let 'innocent' represent an innocent defendant, and let the + symbol indicate the event of a positive litigation outcome for the plaintiff or prosecutor, as the case may be. That is, from the plaintiff or prosecutor's perspective, a positive outcome, or +, occurs when liability is eventually imposed on the defendant. We now proceed to find the values for $\Pr(+|guilty)$, $\Pr(+|innocent)$, $\Pr(guilty)$, $\Pr(innocent)$, and $\Pr(+)$. To begin with, $\Pr(+|guilty)$ is the probability that a guilty defendant will be found guilty at the end of a litigation game. Since we have assumed that the litigation game is 90% sensitive, the value for $\Pr(+|guilty)$ is equal to 0.9. By the same token, $\Pr(+|innocent)$, the probability that a particular litigation game will produce a false positive (i.e., the probability that liability will be imposed on an innocent defendant) is equal to 0.1. This value is 0.1 since, given our initial assumptions, the litigation game produces false positives only 10% of the time.

Now suppose that plaintiffs and prosecutors are risk-averse or virtuous parties, that is, assume that plaintiffs and prosecutors alike are willing to play the litigation game only when they are at least 90% certain that the named defendant has, in fact, committed an unlawful wrongful act.[27] Accordingly, given these stringent assumptions (i.e., risk-averse moving parties and non-random adjudication), the prior probability that a given defendant is guilty is 90%, or stated formally, letting A stand for the prior probability of being guilty, then Pr(A) = Pr(guilty) = 0.9. Summing up, Pr(A) or Pr(guilty) is the prior probability, in the absence of any additional information, that a particular defendant has committed a wrongful act. As stated above, this term is equal to 0.9 since we have assumed that 90% of all named defendants are guilty. Likewise, we determine Pr(B) or Pr(innocent), the prior probability that a particular defendant has not committed any wrongful act. This is simply 1 − Pr(guilty) or 0.1, since 1 − 0.9 = 0.1.

Lastly, Pr(+) refers to the prior probability of a positive litigation outcome — again, 'positive' from the plaintiff's or prosecutor's perspective — in the absence of any information about the defendant's guilt or innocence. This value is found by adding the probability that a true positive result will occur (0.9 × 0.9 = 0.81), plus the probability that a false positive will happen (0.1 × 0.1 = 0.01), and is thus equal to 0.81 plus 0.01 = 0.82. Stated formally, Pr(+) =[Pr(+|guilty) × Pr(guilty)] plus [Pr(+|innocent) × Pr(innocent)]. That is, the prior probability of a positive litigation outcome, Pr(+), is the sum of true positives and false positives and, given our assumptions above, is equal to 0.82 or 82%.

Having translated all the relevant terms of Bayes' theorem, we now restate our Bayesian model of litigation game and find the posterior probability, Pr(guilty|+), that civil or criminal liability will incorrectly imposed on a guilty defendant (i.e., the probability that a defendant who has not committed a wrongful act will nevertheless be incorrectly classified as a wrongful or guilty defendant):

---

[27] This risk-averse conduct is considered 'virtuous' in our model since such moving parties are less willing than their risk-loving colleagues to gamble on the outcome of litigation, or expressed in legal language, virtuous civil plaintiffs rarely file frivolous claims and virtuous criminal prosecutors rarely abuse their discretion. The reader may rest assured, however, that we will relax these unrealistic assumptions later.

Pr(guilty|+) = [Pr(+|guilty) × Pr(guilty)] ÷ Pr(+)

= [Pr(+|guilty) × Pr(guilty)] ÷ ([Pr(+|guilty) × Pr(guilty)] + [Pr(+|innocent) × Pr(innocent)])

= (0.9 × 0.9) ÷ [(0.9)(0.9) + (0.1)(0.1)]

= 0.81 ÷ 0.82 = 0.988

In other words, given our rosy assumptions above, the outcome of any particular litigation game will be highly accurate. Specifically, the probability that a defendant who is found liable for a wrongful act is actually guilty of committing such wrongful act is close to 99%, a value that appears to vindicate Hart and Sacks's optimistic vision of legal process, though there is still a 1% probability that an innocent defendant will nonetheless be found liable. But what happens when the litigation game is played by strategic plaintiffs or zealous prosecutors? That is, what happens when plaintiffs file a greater proportion of frivolous claims (relative to the optimal level of frivolous claims) or when prosecutors routinely 'overcharge' criminal defendants with extraneous or vague offenses (e.g., conspiracy)? We turn to this possibility below.

### 4.2. Non-random adjudication with risk-loving moving parties

Suppose the litigation game is still highly sensitive and specific as before (i.e., 90% sensitive and 90% specific), but that plaintiffs and prosecutors are risk-loving or less-than-virtuous actors. Specifically, assume that the moving parties are willing to play the litigation game even when they are only 60% certain (instead of 90% certain, as we assumed earlier) that the named defendant has committed a wrongful act.[28] The intuition behind this revised assumption is that, in reality, the litigation game might be played by litigants (as well as judges) who are engaged in rent-seeking and self-serving behavior.[29] Thus, with risk-loving moving parties, the prior probability, Pr(guilty), that a given defendant is guilty is now only 60%, while the prior probability, Pr(innocent), that a particular defendant has not committed a wrongful act is 1 − Pr(guilty), or 1 − 0.6 = 0.4. Stated formally: Pr(guilty) = 0.6, and Pr(innocent) = 0.4.

---

[28] Such behavior is 'less-than-virtuous' in our model because the moving party is less concerned with the defendant's actual guilt than a risk-averse or virtuous moving party.

Next, we find the probability that a guilty defendant will be found guilty, or Pr(+|guilty). In this variation of our model, the value for Pr(+|guilty) is equal to 0.90 since we continue to assume the litigation game is 90% sensitive. Pr(+|innocent), the probability that a particular litigation game will produce a false positive (i.e., the probability that liability will be imposed on an innocent defendant), remains 0.1. Lastly, recall that Pr(+) is the probability that a true positive result will occur (in this case, $0.9 \times 0.6 = 0.54$), plus the probability that a false positive will happen ($0.1 \times 0.4 = 0.04$), and is thus equal to 0.54 plus 0.04 = 0.58. Stated formally, Pr(+) = [Pr(+|guilty) × Pr(guilty)] plus [Pr(+|innocent) × Pr(innocent)] = 0.54 plus 0.4 = 0.58.

Given these revised assumptions — non-random adjudication and less-than-virtuous plaintiffs — we now find the posterior probability that liability will be correctly imposed on a guilty or wrongful defendant as follows:

---

[29] For further exploration of this problem, see generally Galanter, 'Why the "Haves" Come Out Ahead' (n 3) and Gordon Tullock, *The Logic of the Law* (n 17). In principle, a more hard-core 'risk-loving' moving party might be willing to gamble on the litigation game even when he or she is only 50% certain of the outcome. Nevertheless, we assume that a risk-loving moving party requires a 60% probability of a positive litigation outcome simply because he or she must expend resources to play the litigation game. Put another way, since the litigation game is not costless — a point made in F.E. Guerra-Pujol, 'Coase's Paradigm' (2011) 1 Indian JLE 1, 27-32; see also Galanter, 'Why the "Haves" Come Out Ahead' (n 3) — and thus, broadly speaking, the higher the cost of playing the litigation game (relative to the resources of the moving party), the more risk-averse an otherwise risk-loving moving party will be.

Pr(guilty|+) = [Pr(+|guilty) × Pr(guilty)] ÷ Pr(+)

= [Pr(+|guilty) × Pr(guilty)] ÷ ([Pr(+|guilty) × Pr(guilty)] + [Pr(+|innocent) × Pr(innocent)])

= (0.9 × 0.6) ÷ [(0.9)(0.6) + (0.1)(0.4)]

= 0.54 ÷ 0.58 = 0.931

In this case, despite the presence of risk-loving moving parties, the outcome of any particular litigation game will still be highly reliable. Specifically, although there is a 7% chance that an innocent defendant will be found liable, the posterior probability that a defendant who is found liable for a wrongful act is actually guilty is still 93%, a value that, once again, appears to affirm the Hart and Sacks vision of the legal system.[30] But now, consider what happens when litigation is a crapshoot, that is, stated formally, what happens when the litigation game is only 50% sensitive and 50% specific?

## 4.3 Random adjudication with risk-averse moving parties

Suppose now that the litigation game is only 50% sensitive and 50% specific. In other words, suppose litigation games are completely random.[31] Under this seemingly unusual scenario, the process of adjudication is no better than a coin toss. Although this assumption may appear fanciful, as we explained earlier,[32] the randomness of adjudication might be a function of the level of the complexity or the level of ambiguity of the applicable legal doctrines (e.g., assumption of risk) or procedural rules (e.g., res judicata). In plain English, the more complex or ambiguous the applicable law is, the more random or arbitrary the outcome of litigation will be.

---

[30] Hart & Sacks, *The Legal Process* (n 26).

[31] To this end, consider the following statement by one of the author's favorite professors in law school: 'Litigation is a crapshoot'. John Langbein, Sterling Professor of Law and Legal History, Yale Law School (New Haven, Conn). The author does not recall the precise date when this proposition was made, but this statement, like Holmes's prediction theory of law, has had a profound influence on our thinking about the legal process. For an empirical exploration of the randomness of litigation, see F.E. Guerra-Pujol, 'Chance and Litigation' (n 25).

[32] See text accompanying notes 24 and 25.

In summary, random adjudication produces purely random results, no better than a coin toss, since it will correctly determine with one-half probability, or p = 0.5, whether the defendant has or has not committed a wrongful act. Given this inherent randomness, along with the presence of virtuous or risk-averse moving parties, we now turn to Bayes' rule to determine the posterior probability that liability will be incorrectly imposed on an innocent defendant (i.e., the probability that a defendant who has not committed a wrongful act will be incorrectly classified as a wrongful or guilty defendant). Again, let 'guilt' stand for a guilty defendant, 'innocent' an innocent defendant, and the symbol + the event of a positive litigation outcome for the moving party (plaintiff or prosecutor). Next, we find the values for Pr(guilt), Pr(innocent), Pr(+|guilt), Pr(+|innocent), and Pr(+).

First, assuming that plaintiffs and prosecutors are virtuous or risk-averse actors and thus are willing to play the litigation game only when they are at least 90% certain that the named defendant is guilty, then Pr(guilty), the prior probability in the absence of other information that a particular defendant has committed a wrongful act, will be equal to 0.9, or stated formally, Pr(guilty) = 0.9. Likewise, Pr(innocent), the prior probability in the absence of other information that a particular defendant has not committed a wrongful act, is simply 1 − Pr(guilty) or 0.1, since 1 − 0.9 = 0.1

Next, Pr(+|guilty), the probability that liability will be imposed on a defendant who is actually guilty, is 0.5 since the litigation game in this variation of our model purely random (i.e., 50% sensitive). Similarly, Pr(+|innocent), the probability that liability will be imposed on an innocent defendant, is also 0.5 since, given our revised assumptions, the litigation game will produce a false positive half of the time the game is played.

Lastly, recall that Pr(+) is the sum of true positives and false positives, that is, the prior probability of a positive litigation outcome, positive from the plaintiff's or prosecutor's perspective, in the absence of any information about the defendant's guilt or innocence. Specifically, given our assumptions above, this value is equal to 0.5, that is, $0.5 \times 0.9 = 0.45$ (true positives) plus $0.5 \times 0.1 = 0.05$ (false positives). Thus, the prior probability of a positive litigation outcome, Pr(+), absent any information about the defendant's guilt or innocence, is equal to 50%.

Thus, given random adjudication and virtuous or risk-averse plaintiffs, we apply Bayes' theorem as follows:

Pr(guilty|+) = [Pr(+|guilty) × Pr(guilty)] ÷ Pr(+)
= [Pr(+|guilty) × Pr(guilty)] ÷ ([Pr(+|guilty) × Pr(guilty)] + [Pr(+|innocent) × Pr(innocent)])
= (0.5 × 0.9) ÷ [(0.5)(0.9) + (0.5)(0.1)]
= 0.45 ÷ 0.50 = 0.9

This result is perhaps the most surprising one thus far. Even when the litigation game is a purely random process, no better than a coin toss, the outcome of any individual litigation game will still be highly reliable, given the presence of virtuous moving parties. Specifically, under this scenario there is a 90% probability that a defendant who is found liable for a wrongful act is, in fact, actually guilty.[33] Although this value is less than the corresponding values for Pr(guilty|+) in the previous two permutations of the model (subsections 4.1 and 4.2 above), this difference is marginal at best, considering the enormous qualitative differences between non-random adjudication and a purely random legal system. The present permutation of the model, however, assumes the presence of virtuous plaintiffs and prosecutors. What happens when the litigation game is purely random and the moving parties are less-than-virtuous? We explore this intriguing possibility in subsection 4.4 below.

---

[33] In other words, even when the outcome of litigation is random, there is only a 10% chance that an innocent defendant will be found guilty or civilly or criminally liable.

## 4.4 Random adjudication with risk-loving moving parties

Now suppose the litigation game is still a crapshoot but that plaintiffs and prosecutors are risk-loving or 'less-than-virtuous'; that is, assume that the moving parties are more willing to gamble than their virtuous colleagues. Specifically, we will assume that the litigation game is 50% sensitive and 50% specific and that plaintiffs and prosecutors are willing to play the litigation game even when they are only 60% certain that the named defendant has committed a wrongful act. Although these assumptions do not appear to be plausible, this permutation of our model, however implausible, may nevertheless provide an instructive counter-factual or hypothetical illustration of our Bayesian approach to litigation.[34]

Given our revised assumptions (i.e., random results and risk-loving or less than virtuous actors), we once again turn to Bayes' theorem to determine the posterior probability that liability will be incorrectly imposed on an innocent defendant (i.e., the probability that a defendant who has not committed a wrongful act will be incorrectly classified as a wrongful or guilty defendant), and once again, 'guilt' stands for a guilty defendant, 'innocent' indicates an innocent defendant, and the symbol + represents the event of a positive litigation outcome for the plaintiff or prosecutor.

As such, in the absence of any additional information or evidence, Pr(guilty), the prior probability that a particular defendant has committed a wrongful act, is equal to 0.6, while Pr(innocent), the prior probability that a particular defendant has not committed a wrongful act, is 0.4 (i.e., 1 − Pr(guilty), or 1 − 0.6). Next, Pr(+|guilty), the probability that liability will be imposed on a defendant who is actually guilty, and Pr(+|innocent), the probability that liability will be imposed on an innocent defendant, are both equal to 0.5 since, given our assumptions, this version of the litigation game is purely random. Lastly, Pr(+),the sum of true positives and false positives, is also 0.5 since, given our assumptions above, 0.5 × 0.6 = 0.3 (true positives) and 0.5 × 0.4 = 0.2 (false positives), or put another way, the prior probability of a positive litigation outcome (again, from the plaintiff's or prosecutor's perspective), absent any information about the defendant's guilt or innocence, is equal to 50%.

---

[34] This scenario, however, would be plausible in the presence of risk-loving actors, or if we picture the litigants as pure gamblers.

Therefore, given random adjudication and risk-loving plaintiffs, we now apply Bayes' theorem as follows:

Pr(guilty|+) = [Pr(+|guilty) × Pr(guilty)] ÷ Pr(+)
= [Pr(+|guilty) × Pr(guilty)] ÷ ([Pr(+|guilty) × Pr(guilty)] + [Pr(+|innocent) × Pr(innocent)])
= (0.5 × 0.6) ÷ [(0.5)(0.6) + (0.5)(0.4)]
= 0.3 ÷ by 0.5 = 0.6

What is most surprising about this result is the ability of the litigation process to produce reliable results more than half the time, even when the underlying litigation game itself is purely random and even when the actors are less than virtuous. Specifically, the probability that the outcome of any individual litigation game will be accurate is 60%, even though the underlying litigation game is purely random, no more reliable than a coin toss. One way of explaining this potential paradox is to take another look at the Pr(guilty) term: the prior probability in the absence of additional information that a defendant selected at random is guilty (i.e., the prior probability that a particular defendant has committed a wrongful act). This prior probability term exerts a decisive influence in the fourth permutation of our model precisely because the outcome of litigation is purely random. That is, when litigation is a crap shoot, or to be more precise, when litigation is a coin toss, both the prior and posterior probabilities of the defendant's guilt are the same. Here, since Pr(guilt) = 0.6, then Pr(+|guilty) = 0.6.

## 5. Conclusion

We wish to close this paper with a confession. Ex ante, before researching and writing this paper, we took a dim view of the litigation game. Given the complexity and ambiguity of substantive as well as procedural rules, the indeterminate nature of most legal standards, and the high levels of strategic behavior by both litigants and judges, we expected our Bayesian model to confirm this negative view of the legal process. Ironically, however, the results of our Bayesian model of the litigation game are still surprising. In essence, they show that, regardless of the operative rules of procedure and substantive legal doctrine, 'positive' litigation outcomes (as defined in this paper) are nevertheless a highly reliable indicator of a defendant's guilt. Specifically, our model demonstrates that when a defendant is found guilty of committing a wrongful act (civil or criminal), there is a high posterior probability that the defendant actually committed such wrongful act, even when the underlying process of adjudication is random and even when the moving parties are risk-loving or less-than-virtuous.